# Lattice thermal conductivity and mechanical properties of the single-layer penta-NiN$_2$ explored by a deep-learning interatomic potential

Pedram Mirchi[1], Christophe Adessi[2], Samy Merabia[2] and Ali Rajabpour [1,2,3,*]

[1] Advanced Simulation and Computing Laboratory (ASCL), Mechanical Engineering Department, Imam Khomeini International University, Qazvin, Iran

[2] Univ Lyon, Univ Claude Bernard Lyon 1, CNRS, Institut Lumière Matière, F-69622, Villeurbanne, France

[3] School of Nano Science, Institute for Research in Fundamental Sciences (IPM), 19395-5531, Tehran, Iran

**Abstract**

Penta-NiN$_2$, a novel pentagonal 2D sheet with potential nanoelectronic applications, is investigated in terms of its lattice thermal conductivity, stability, and mechanical behavior. A deep learning interatomic potential (DLP) is firstly generated from *ab-initio* molecular dynamics (AIMD) data and then utilized for classical molecular dynamics simulations. The DLP's accuracy is verified, showing strong agreement with AIMD results. The dependence of thermal conductivity on size, temperature, and tensile strain, reveals important insights into the material's thermal properties. Additionally, the mechanical response of penta-NiN$_2$ under uniaxial loading is examined, yielding a Young's modulus of approximately 368 GPa. The influence of vacancy defects on mechanical properties is analyzed, demonstrating significant reduction in modulus, fracture stress, and ultimate strength. This study also investigates the influence of strain on phonon dispersion relations and phonon group velocity in penta-NiN$_2$, shedding light on how alterations in the atomic lattice affect the phonon dynamics and, consequently, impact the thermal conductivity. This investigation showcases the ability of deep learning based interatomic potentials in studying the properties of 2D Penta-NiN$_2$.

## 1. Introduction

Following the development of graphene, researchers have become intrigued by other two-dimensional materials exhibiting diverse mechanical, thermal, optical, and electrical properties. 2D nanostructures can display a wide spectrum of thermal transport properties, spanning from exceptionally low to extremely high conductivity. Moreover, their mechanical and thermal properties can be finely adjusted [1-6]. Although numerous two-dimensional materials made of nitrogen atoms have been created so far, attempts to synthesize nitrogen-rich structures have been less successful. Recently, Bykov et al. fabricated a two-dimensional structure of beryllium polynitride, denoted as BeN4, utilizing a high-pressure synthesis technique [7]. Ghorbani et al. examined the anisotropic mechanical and thermal properties of monolayers consisting of $BeN_4$, $MgN_4$, and $PtN_4$ [8]. Mortazavi et al. conducted an investigation into the inherent physical properties and stability of diverse pentagonal metal diazenide nanomembranes [9]. One of these materials, Penta-$NiN_2$, represents a novel pentagonal 2D sheet that was recently synthesized using high-pressure techniques and crystal-chemical design [10]. Displaying characteristics of a direct band gap semiconductor, Penta-$NiN_2$ holds promise for applications in nanoelectronics. This atomic-thick planar sheet consists of $Ni_2N_3$ pentagons arranged in a Cairo-type mosaic pattern [11]. Zhang et al. investigated the intrinsic lattice thermal conductivity of Penta-$NiN_2$, along with its thermal behavior under biaxial tensile strain. Their approach involved solving the Boltzmann transport equation (BTE) and incorporating four-phonon scattering mechanisms in conjunction with the self-consistent phonon (SCPH) theory [12]. Zhang et al. employed the CellMatch package to systematically explore the phase space of layer combinations, resulting in the generation of twisted structures for bilayer penta-$NiN_2$ [13]. Li et al. conducted research on the adsorption of small molecules on a penta-$NiN_2$ nanosheet and investigated its impact on electronic characteristics[14].

---

[*] Corresponding Authors: A.Rajabpour, Email: rajabpour@eng.ikiu.ac.ir

The fundamental methods for computing the properties of 2D materials, including mechanical properties and thermal conductivity, involve density functional theory combined with the Boltzmann transport equation (DFT-BTE), as well as classical molecular dynamics (MD) simulations. Within the MD methodology, the accuracy of the results is notably influenced by the quality of the interatomic potential. Different interatomic potentials can lead to various estimations of thermal and mechanical quantities. For instance, while the range of values of graphene thermal conductivity spans from 1500 to 5300 W/mK [15,16], MD simulations predict a range of about 350 to 3000 W/mK [17,18]. Furthermore, in the application of DFT-BTE approaches, the outputs depend on the choice of the exchange-correlation functional and the computational recipes for calculating 2nd and 3rd-order force constants [19,20]. By utilizing machine learning interatomic potentials, it becomes feasible to strike a balance between classical and *ab initio* molecular dynamics (AIMD). Various machine learning models have been proposed in the literature, including the Behler-Parrinello neural network potentials (BPNNP) [21], Gaussian approximation potentials (GAP) [22], Spectral Neighbor Analysis Potential (SNAP) [23], Moment Tensor Potentials (MTP) [24], ANI-1 [25], and SchNet [26].

One of the most widely adopted approaches for computing interatomic potentials is Deep Potential Molecular Dynamics (DeePMD) package [27]. DeePMD offers an advantage over other neural network approaches like BPNN, as they eliminate the need for manual construction of local symmetry functions [28]. DeePMD models have proven to be highly effective in investigating thermal and mechanical properties. Several studies have been undertaken using DeePMD to compute thermal conductivity in liquids [29–31], semiconductors [32], and 2D nanostructures such as SnSe [33], GeS, and SnS [34]. Feng et al. explored the molten $LaCl_3$ structure using DeePMD [35], while Sours et al. predicted the mechanical and structural properties of pure silica zeolites [36]. Matusalema et al. employed AIMD computations and deep learning techniques to scrutinize the plastic deformation behavior of high-pressure, high-temperature water ices [37]. Additionally, Du et al. predicted melting points and elastic constants using the deep potential model [38]. In order to explore the segregation of W into ZrB2 grain boundaries and the strengthening effect on grain boundaries induced by segregation at high temperatures, Dai et al. developed a deep learning potential for the $Zr_{1-x}W_xB_2$ system [39]. Furthermore, Wang et al. formulated a deep neural network potential to simulate the structural features of the formation of diverse forms of carbon [40].

In this study, we generate and employ a deep learning-based interatomic potential for conducting molecular dynamics simulations, enabling an in-depth analysis of the mechanical properties and thermal conductivity of penta-$NiN_2$ monolayers. In order to construct a deep learning interatomic potential (DLP), the data set is first generated using AIMD trajectories before employing the deep learning technique to train the interatomic potential. We exploit this trained interatomic potential to determine both thermal and mechanical properties of penta-$NiN_2$ using MD simulations. On this occasion, we investigate the effect of strain on thermal conductivity and the effect of vacancies on mechanical properties.

## 2. Simulation Details

This section presents the methodology employed to extract deep learning-based interatomic potential from AIMD. Furthermore, the approach of applying these potentials in classical simulations to evaluate the mechanical properties and thermal conductivity of penta-$NiN_2$ will be discussed.

*2.1 Ab-initio simulation*

In this study, we employed the GGA/PBE functional and the Vienna *Ab initio* Simulation Package (VASP) to conduct AIMD simulations. The structure was optimized under various strains (0-12%) and temperatures (100-600 K). A plane-wave cutoff energy of 500 eV was selected for the structures. Furthermore, simulations were carried out using 20,532 time steps of 0.5 fs each, considering 5×4×1 supercells containing 120 atoms.

*2.2 Interatomic potential training*

The DeePMD-kit was employed to train the deep learning potential. In contrast to other machine learning techniques, deep learning demonstrated remarkable performance in handling extensive datasets from AIMD, particularly those encompassing complex and symmetry-invariant properties [27]. The foundational principle of this model lies in the notion that the energy of a system containing *N* atoms can be computationally deduced from the characteristics of individual atoms. This principle can be expressed as follows:

$$E = \sum_i E_i \quad (1)$$

where the energy of atom *i* denoted as $E_i$, is intricately linked with its immediate local atomic arrangement. To precisely define this local configuration for atom *i*, the descriptor matrix $D_{ij}$ is employed for atom *j* ensuring that atom *j* is situated within the designated cutoff radius relative to atom *i*.

$$D_{ij} = \{\frac{1}{R_{ij}}, \frac{x_{ij}}{R_{ij}}, \frac{y_{ij}}{R_{ij}}, \frac{z_{ij}}{R_{ij}}\} \quad (2)$$

where $x_{ij}$, $y_{ij}$, and $z_{ij}$ are the relative coordinates, while $R_{ij}$ represents the distance between *i* and *j*. Moreover, the atomic energy can be written as

$$E_i = \lambda_{(i)}(D_{ij}) \tag{3}$$

where $\lambda$ represents the multilayer perceptron with the hidden layer. During the simulation, a cutoff radius of 6 Å was imposed and the loss function $L$ is defined as

$$L(p_\epsilon, p_f, p_\xi) = \frac{p_\epsilon}{N}\Delta E^2 + \frac{p_f}{3N}\sum_i |\Delta F_i|^2 + \frac{p_\xi}{9N}\sum_i \|\Delta v\|^2 \tag{4}$$

Here, $N$ represents the total number of atoms, while $p_\epsilon$, $p_f$ and $p_\xi$ denote adjustable coefficients. $\Delta E$, $\Delta F_i$ and $\Delta v$ stand for the root mean square (RMS) errors associated with energy, force, and virial stress, respectively. |...| and ||...|| represent the norm of the force vector and stress tensor. Throughout the training phase, the embedding network consisted of {25, 50, 100} layers, and the fitting network had {240, 240, 240} layers. The training process spanned 2 million epochs, with the learning rate gradually decreasing exponentially from $1 \times 10^{-3}$ to $3.51 \times 10^{-8}$. Once the DLP was generated, classical molecular dynamics simulations were conducted.

*2.3 Molecular dynamic simulations*

To forecast the mechanical properties and lattice thermal conductivity of the penta-NiN$_2$ monolayer, we employed non-equilibrium molecular dynamics (NEMD) simulations [41–44]. The simulations were conducted using the Large-scale Atomic/Molecular Massively Parallel Simulator (LAMMPS) software package [45], which incorporated the DLP potential for precise atomic interactions. The analysis of the penta-NiN$_2$ nanostructure was performed under periodic boundary conditions, aiming to mitigate finite-length effects and the influence of boundary atoms. The simulations were conducted at an average temperature of 300 K, employing a time step of 1 fs. Thermal conductivity calculations can be performed using non-equilibrium molecular dynamics (NEMD) techniques. Multiple approaches exist for calculating thermal conductivity using the NEMD technique. The first method involves creating two distinct regions, wherein energy is injected into the hot zone and extracted from the cold zone to maintain their respective temperatures. This approach enables the determination of thermal conductivity by assessing energy transfer and a steady temperature gradient between these two regions [46,47]. An alternative technique for calculating thermal conductivity via NEMD includes the controlled addition and subtraction of a specific amount of energy between two distinct regions. This method facilitates the determination of thermal conductivity based on the energy transfer and temperature gradient observed across these two regions [48]. Furthermore, an additional approach, known as reverse non-equilibrium molecular dynamics (RNEMD), maybe devised utilizing an algorithm proposed by Müller-Plathe [49]. In this approach, a kinetic energy swap is applied to create a temperature gradient within the sample, enabling the investigation of its thermal behavior.

The thermal conductivity ($\kappa$) is determined using Fourier's law, which is expressed as follows:

$$\kappa = -q'' / \frac{dT}{dx} \tag{5}$$

Here, $q''$ represents the heat flux flowing through the structure, and $\frac{dT}{dx}$ represents the temperature gradient across the structure.

Numerous research studies have employed either Cauchy stress [50,51] or virial stress [52,53] for stress analysis. Cauchy stress offers computational efficiency advantages; however, it tends to introduce non-physical initial stress at elevated temperatures. In contrast, virial stress ensures a zero initial stress state. Hence, for our study, we have utilized virial stress. The values of the stress are recorded as follows [54],

$$S = \frac{1}{V} \sum_{a \epsilon V} \left[ -m\vec{v}_a \otimes \vec{v}_a + \frac{1}{2} \sum_{a \neq b} (\vec{r}_{ab} \otimes \vec{F}_{ab}) \right] \tag{6}$$

where, $S$ denotes the stress tensor, $V$ represents the volume of the structure. $m$ and $\vec{v}_a$ refer to the mass and velocity vector, respectively. The position and force vectors between atoms a and b are indicated by $\vec{r}_{ab}$ and $\vec{F}_{ab}$, and the outer product is symbolized by $\otimes$. The workflow for training and customizing DLP models is illustrated in Figure 1.

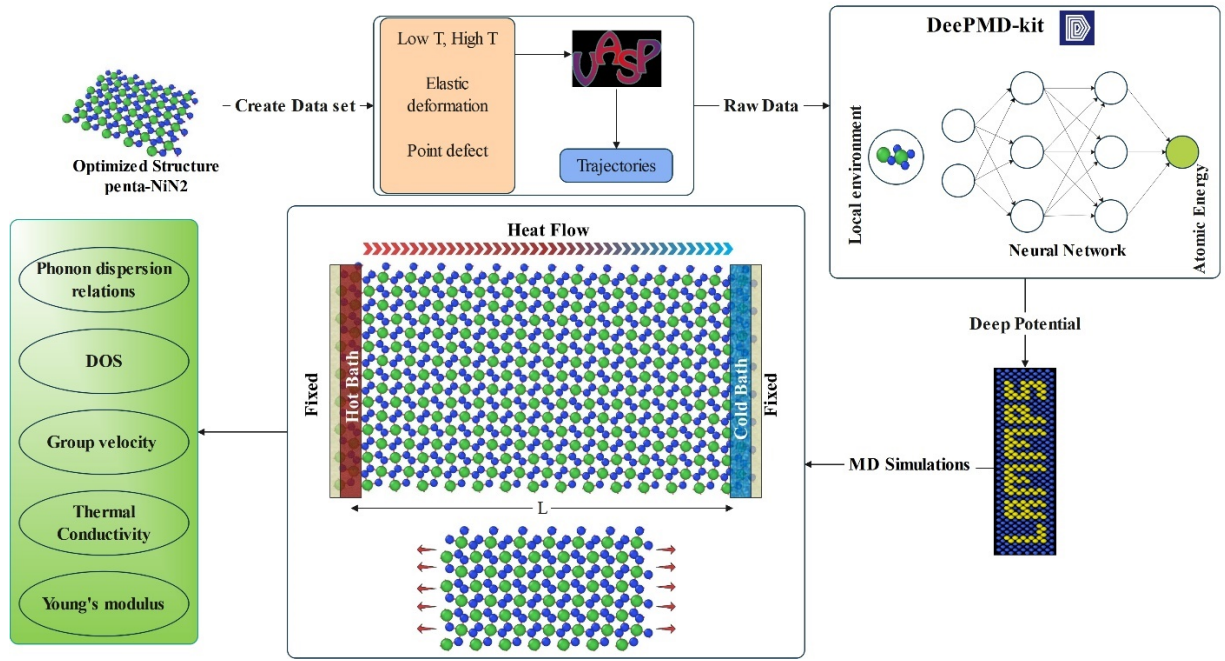

Figure 1. Training a Deep Learning Potential (DLP) using AIMD trajectories to predict thermal and mechanical properties. DOS means the density of states.

## 3. Results and discussion

The main objective of this study is to develop a DLP capable of accurately predicting both the thermal conductivity and mechanical properties of the penta-$NiN_2$ monolayer using classical molecular dynamics. This necessitates the creation of a comprehensive training dataset that encompasses the full range of configurations encountered during classical MD simulations. As a result, AIMD simulations are performed across a range of temperatures from 100 K to 600 K, as well as for different strains ranging from 0% to 12%, to generate the requested training datasets.

Figure 2a illustrates the crystal structure of penta-$NiN_2$, showing evidence of a flat arrangement characterized by P4/mbm symmetry. The unit cell contains two Ni atoms and four N atoms. Our calculations yield lattice parameters of 4.53 Å along both the x and y directions. The lengths of Ni-N and N-N bonds are determined to be 1.88 Å and 1.24 Å, respectively, in agreement with previous computational findings [11,12]. Moving to Figure 2b, a comparison is presented between the phonon dispersion relations (PDR) and the phonon Density of States (DOS) as predicted by both DLP and Density Functional Perturbation Theory (DFPT). As can be observed, there is a good agreement between the phonon dispersion relation and the DOS estimated using DLP and DFPT. It is worth noting that no imaginary frequency is observed which confirms the reliability of quantum and DLP calculations.

Furthermore, our investigation explores the predictive capabilities of the DLP developed. By harnessing the comprehensive training dataset, we examine the DLP's performance in estimating thermal conductivity and mechanical properties, thereby revealing its effectiveness in bridging the gap between computational efficiency and accuracy. These findings provide valuable insights into the potential of DLPs as powerful tools for materials research and property predictions, potentially improving the way we approach material characterization and design.

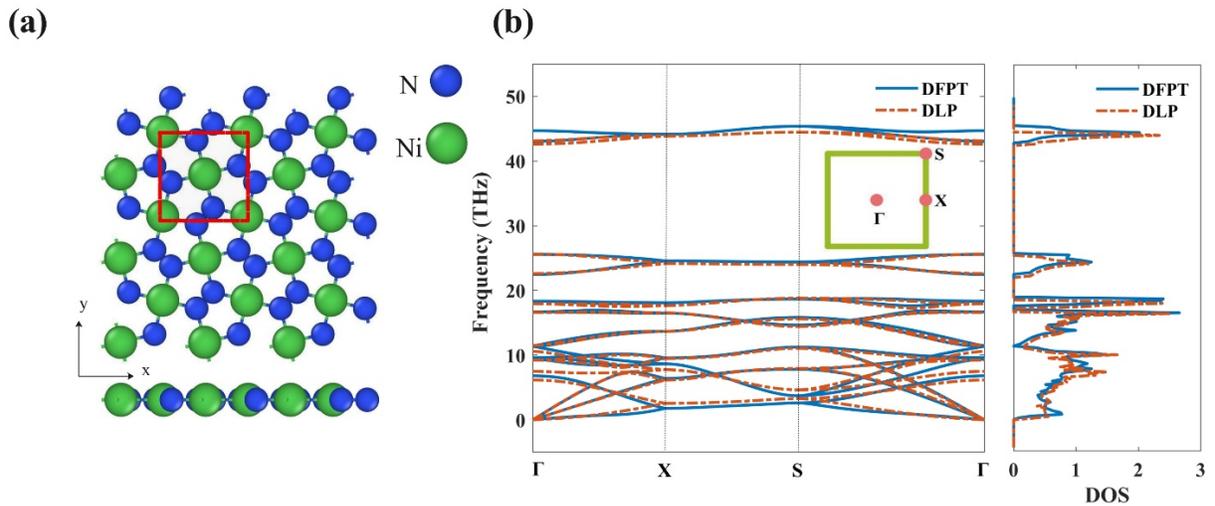

Figure 2. (a) Geometric structure depicted from both top (top) and side (bottom) perspectives (b) Phonon dispersion relations and phonon density of states for Penta-NiN$_2$.

The accuracy of the trained DLP was assessed utilizing a testing dataset encompassing 350 configurations. This dataset comprises varying temperatures and levels of stress, enabling a comprehensive evaluation of the potential model's performance across different conditions. Figure 3 illustrates the correlation between the energy per atom, force, and virial stress per atom as predicted by both AIMD and the DLP.

The root-mean-squared errors (RMSE) for energy per atom, force, and virial stress per atom are remarkably low, implying a substantial level of matching between the DLP predictions and the AIMD outputs. Furthermore, the exceptionally high correlation coefficient of 0.99 provides additional support for the very good accuracy of DLP, reinforcing the conclusion that the trained DLP closely mirrors the accuracy achieved through AIMD with a reduced computational cost. Moreover, the phonon group velocities calculated from the trained MD potential and from the DFPT modeling have been calculated and compared, as demonstrated in the appendix (Figure S1). The obtained results indicate a substantial agreement between the two methods.

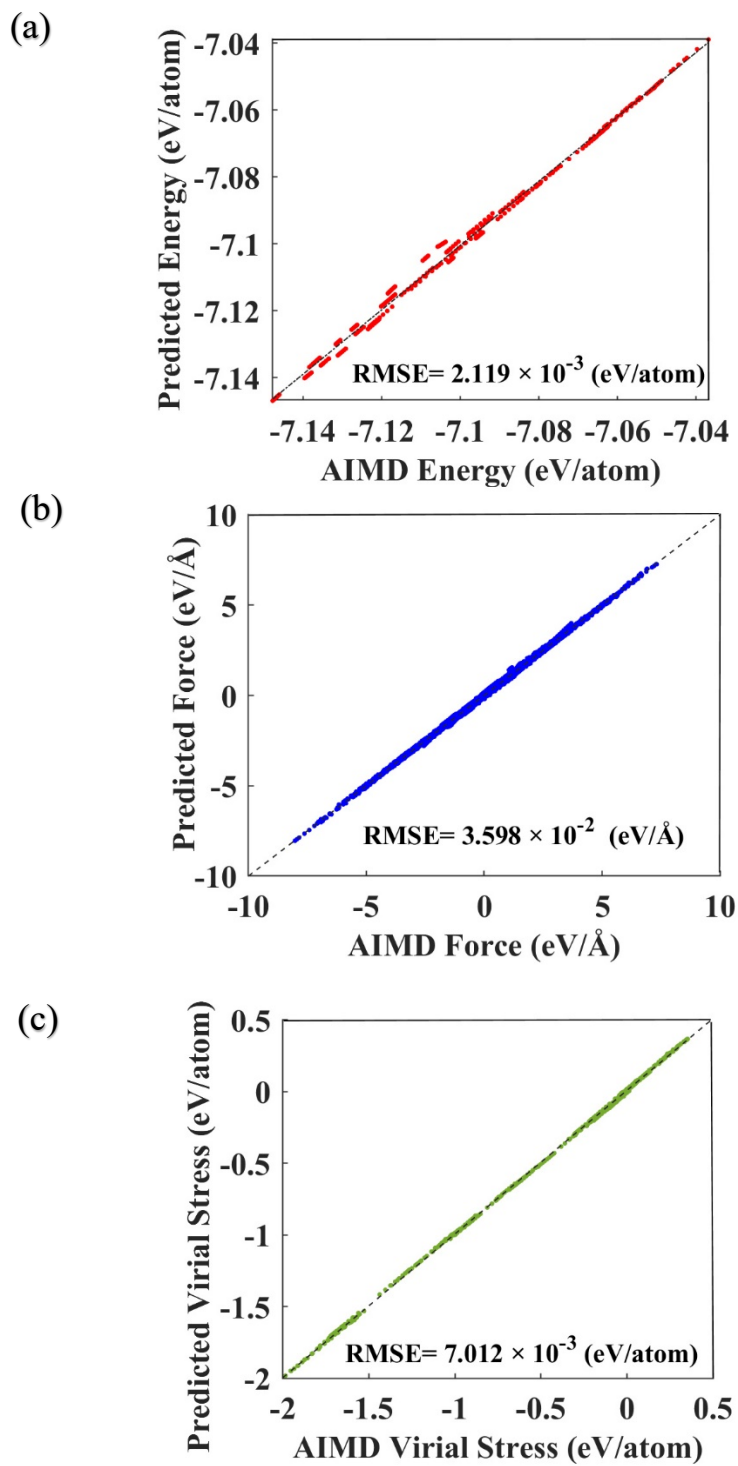

Figure 3. Correlations between the data from AIMD and DLP. (a) energy, (b) force magnitude, and (c) virial stress magnitude.

Non-equilibrium molecular dynamics approach is employed to compute the thermal conductivity. Figure 4 presents the dynamics of energy fluctuations within both the heated and cooled domains over time. As time advances, the system reaches steady state, where the net energy exchange between these two regions becomes constant. Furthermore, Figure 4 provides an illustration of the temperature distribution throughout the structure. By mitigating nonlinear temperature fluctuations in the vicinity of the heated and cooled regions, a linear temperature gradient is established along the system.

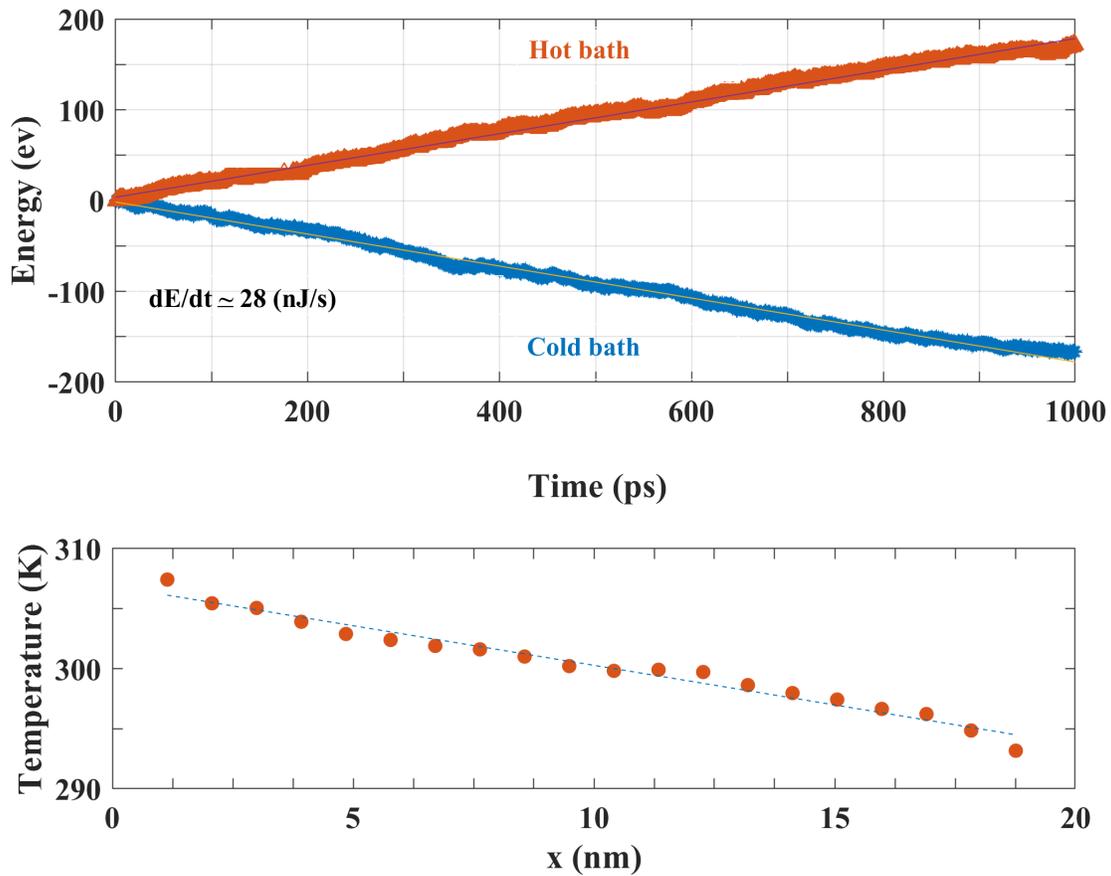

Figure 4. Changes in accumulated energy (top) and the temperature profile of penta-$NiN_2$ (bottom) in the steady state in a non-equilibrium situation.

*3.1 Thermal properties*

In this subsection, we calculate the lattice thermal conductivity of penta-NiN$_2$, employing non-equilibrium molecular dynamics (NEMD) simulations. These simulations are conducted utilizing the DLP. Figure 5 illustrates the impact of length on the thermal conductivity of the penta-NiN$_2$ monolayer at room temperature.

Th size dependent thermal conductivity has been already observed in one-dimensional (1D) materials, such as carbon nanotubes and semiconductor nanowires, as well as in two-dimensional (2D) nanomaterials like graphene [55]. Typically, as the length increases, the thermal conductivity of 2D nanomaterials tends to converge toward a constant value. The thermal conductivity of bulk 2D materials, along with its effective phonon mean free path, can be estimated using the following equation [56].

$$\frac{1}{k_l} = \frac{1}{k_\infty}\left(1 + \frac{\lambda}{L}\right) \qquad (7)$$

In this equation, $k_\infty$ represents the thermal conductivity of the material having an infinite length, $\lambda$ denotes the effective phonon mean free path (MFP) of the material, and $L$ represents the finite length of the sample.

Figure 5 illustrates the relationship between thermal conductivity and length. The plot clearly indicates that the thermal conductivity increases while the length increases particularly for short length. Nevertheless, this dependency wanes as length continues to extend, resulting in the convergence of thermal conductivity at a specific value for longer lengths. However, the outcomes depicted in this figure diverge from the findings of reference [9]. The primary reason for this discrepancy likely arises from the utilization of a broader scope of data in this study and the enhanced precision offered by the deep neural network, in comparison to alternative machine learning methods. By contrast, taking into account the inverse correlation between thermal conductivity and thickness, the results of this study bear a relatively robust correspondence with the findings in Ref. [12].

Moving on, Figure 6 offers an illustration of the thermal conductivity corresponding to a 20nm length of penta-NiN$_2$, capturing its temperature dependence in the range from 100 K to 600 K. In this temperature range, it is anticipated that the thermal conductivity of a crystalline solid material will exhibit an inverse proportionality to temperature [57]. In other terms, when phonon-phonon scattering constitutes the primary contribution to thermal resistance, the thermal conductivity (κ) follows the trend of $k \sim T^{-1}$.

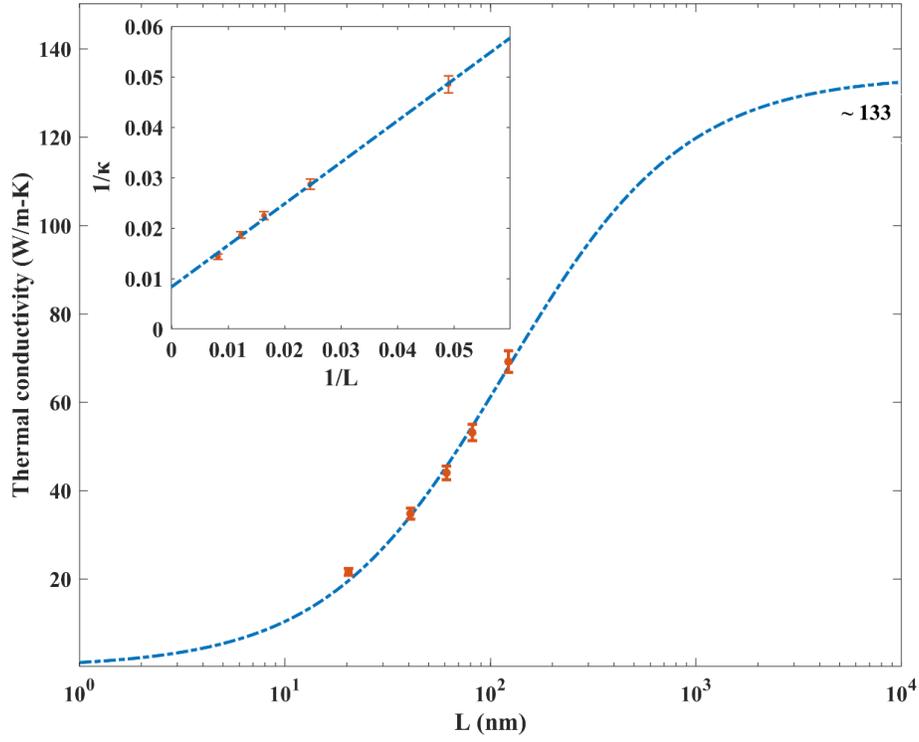

Figure 5. The lattice thermal conductivity of penta-NiN$_2$ as a function of length at 300 K

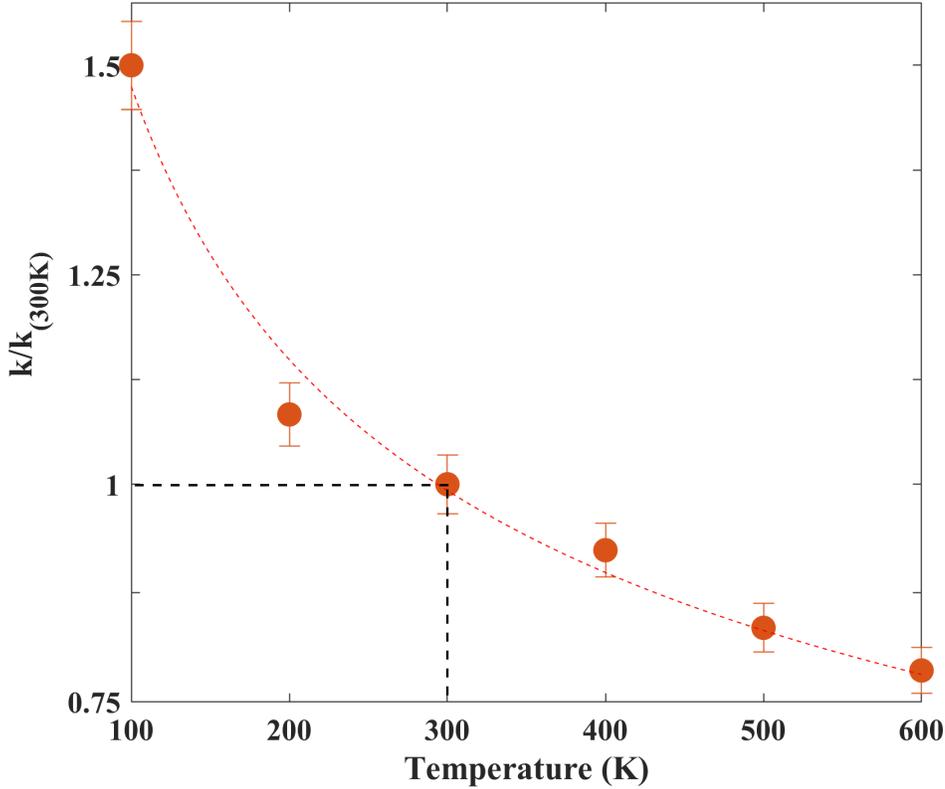

Figure 6. Temperature-dependent lattice thermal conductivity of penta-NiN$_2$ normalized by its thermal conductivity at 300 K.

We also investigate the effect of tensile strain on the thermal conductivity of penta-NiN$_2$. Previous studies have explored the impact of strain on the thermal conductivity of 2D monolayers [58–60]. However, less is known about the influence of tensile strain on the thermal conductivity of penta-NiN$_2$ sheets. To address this point, we apply uniaxial tensile strain in the direction of the heat flux. Figure 7 illustrates the calculated thermal conductivities of penta-NiN$_2$ under different strains. The effect of applied strain on the group velocity and phonon dispersion relation is also shown in Figures S2 and Figures S3 in the supporting information respectively. It is observed that the group velocity at the Gamma point decreases when the strain level is increased, which leads to reduce the thermal conductivity. In other words, as the material is strained, the propagation of phonons becomes slower, impacting the thermal conductivity of the material. This phenomenon can be understood by considering how strain affects the atomic vibrations in the crystal lattice. When a material is strained, the distances between atoms change, altering the interatomic forces. These changes in interatomic forces affect the phonon dispersion, which describes the relationship between the frequency and momentum of phonons in the material. Close to Gamma point, which corresponds to zero momentum, an increase in strain levels causes a decrease in the group velocity of phonons. Group velocity represents the speed at which a wave packet (phonon in this case) propagates through the material. Therefore, this observation reported in Figure S2 suggests that increasing strain levels can be used as a strategy to effectively manipulate the thermal conductivity of materials. By controlling the strain, researchers can modulate the movement of phonons, and consequently, control heat transfer properties in various applications such as thermoelectric materials or thermal insulators.

In this study, two MD potential versions were used to calculate the change in thermal conductivity due to strain. One of the potentials was obtained based on the dataset obtained from quantum calculations considering strained configurations, and the other was calculated based on the dataset in the absence of applied strain. Figure 7 clearly shows that these two versions yield different results for the reduction of the thermal conductivity with strain. Furthermore, it should be noted that the potential trained without strain was unable to maintain structural stability at high strain levels. This indicates that when studying thermal conductivity with strain, it is crucial to employ a potential that is trained on strained data obtained from AIMD calculations.

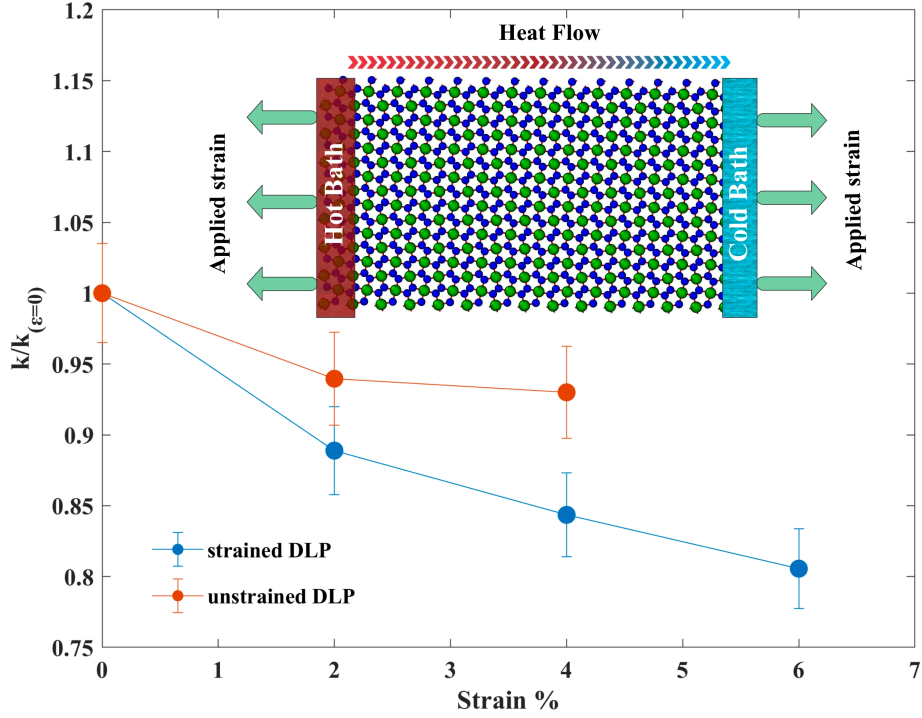

Figure 7. Thermal conductivity at different tensile strains for penta-NiN$_2$ sheet having 20 nm length. The Y-axis values are normalized by the thermal conductivity of the pristine (unstrained) penta-NiN$_2$. The blue points are calculated based on the potential trained on the strained AIMD dataset while the orange ones are based on unstrained dataset.

In order to have a deeper understanding of the effect of strain on thermal conductivity, we also calculate the thermal conductivity in two states, without strain and with applied strain, directly from *ab initio* calculations. All the *ab initio* calculations are performed using the density functional theory (DFT) package SIESTA [61] under the generalized gradient approximation of Perdew, Burke & Ernzerhof [62] and using Troullier–Martin norm-conserving pseudopotentials [63]. The orbital basis used for all the elements is a double zeta polarized one. We have used a 1000 Ry mesh cut off in order to ensure force convergency down to $10^{-4}$ eV/Å and a hydrostatic pressure threshold of 100 bar. For all the calculations a 15x1x15 k-points Monkhorst-pack grid has been used. The harmonic force constants have been computed using the Phonopy package [64] using 5x1x5 supercells.

The thermal conductance $G$ is calculated as:

$$G = -\frac{k_B^2 \cdot T}{h} \int_0^\infty \left(\frac{h\omega}{2\pi k_B T}\right)^2 T(\omega) \frac{\partial f_{BE}}{\partial \omega} d\omega \qquad (7)$$

where $k_B$ is the Boltzmann constant, $T$ the temperature, $h$ the Planck constant, $T(\omega)$ the density of modes (DOM), $f_{BE}$ the Bose-Einstein distribution and $\omega$ the angular frequency. It should be noted that the DOM does not include scattering processes and is derived only from harmonic force constants. Figure 8 shows the thermal conductance variation as a function of temperature in two forms of pristine and strained $NiN_2$. It is shown that at room temperature, the deviation of the thermal conductance (within the harmonic approximation) for a 2 % strain (as compared to the relaxed pristine structure) is 13 % and for a 4 % strain the deviation is 23 %. Thus, it can be seen that *ab initio* calculations also confirm the decreasing effect of strain on the thermal conductivity of $NiN_2$. Due to the fact that in *ab initio* calculations phonon scattering has not been considered and only the harmonic force constants has been taken into account, it can be concluded that the strain has a direct effect on the interatomic forces, which is the main reason to reduce the thermal conductivity of the $NiN_2$ 2D material.

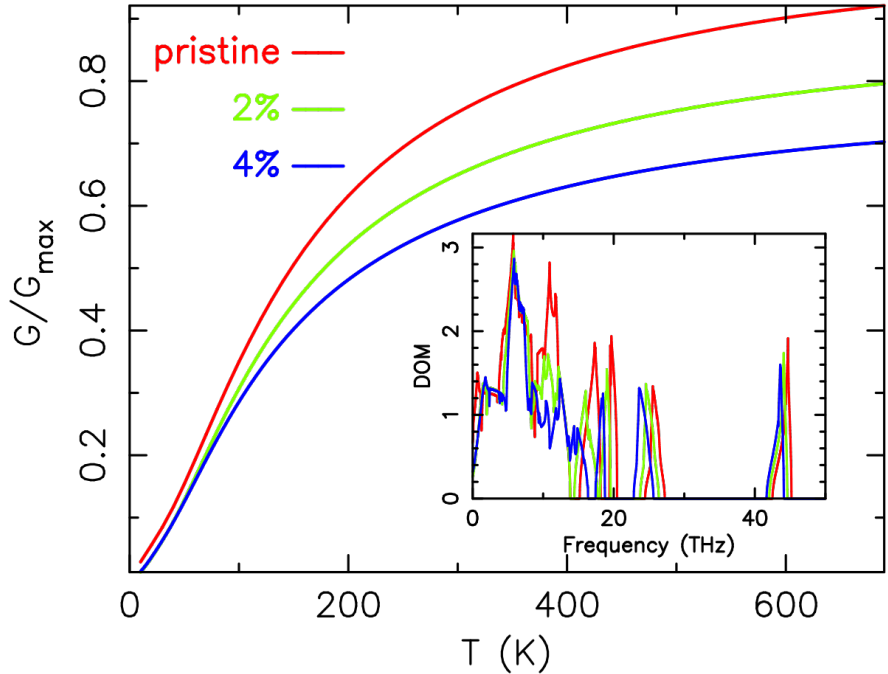

Figure 8: Harmonic thermal conductance (normalized by the maximum value for the pristine system) as a function of the temperature. In the inset, the density of modes (DOM) is given. The green and blue lines correspond to a uniaxial strain of 2 % and 4 % respectively and the red one (denoted pristine) is the unstrained system.

*3.2 Mechanical properties*

Figure 9 illustrates the stress-strain response of the material when subjected to uniaxial loading. The resulting calculations yielded Young's modulus of approximately 368 GPa with a thickness of 0.318 nm. It is important to note that a correlation exists between Young's modulus and thickness, with a decrease in thickness leading to an increase in Young's modulus. Considering this inverse correlation, the Young's modulus value derived from the simulations aligns with the range of results presented in Refs. [11,12] and closely resembles the average of the values reported in Ref. [10]. Figure 9 also highlights that the presence of vacancy defects in the material leads to a decrease in its mechanical properties. In particular, when the material contains 2% defects (single vacancies), there is a significant reduction in its elastic modulus, fractural strain, and ultimate strength. These properties experience reductions of 38%, 29%, and 45% respectively when compared to defect-free conditions.

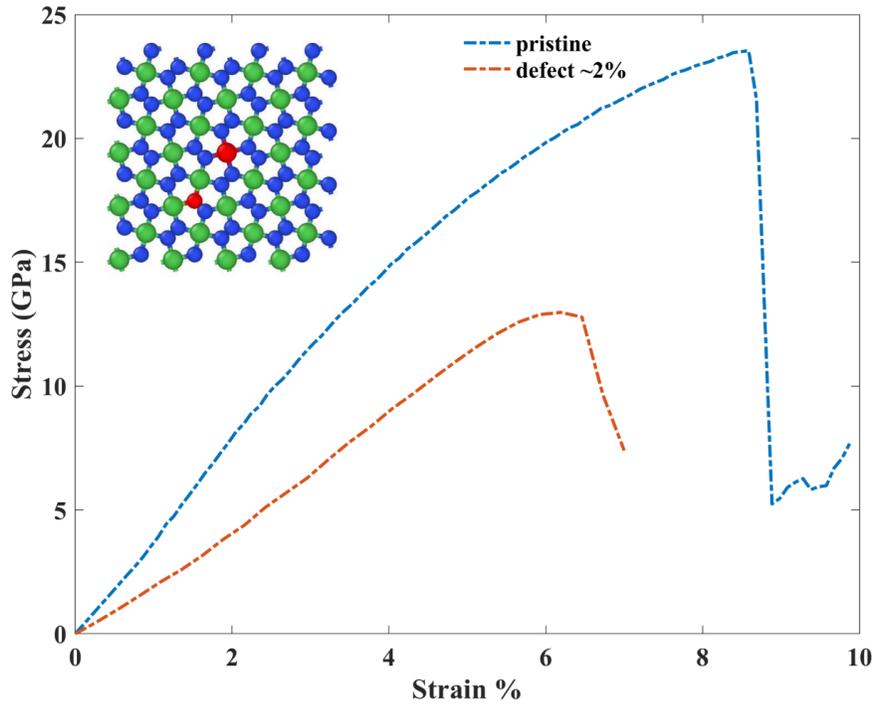

Figure 9. Stress-strain curve under uniaxial tensile load at 300 K for the 5 nm length pristine and defective penta-$NiN_2$ sheets (~ 2% vacancy).

In Table 1 we compare the current results with previous research. It is shown that the thermal conductivity of the NiN$_2$ structure extrapolated at infinite length (based on the fitted function in Figure 5) has been obtained using DLP equal to 133 W/mK, which is within the range of the result predicted by the DFT method employed in Ref.[12]. However, the results of Ref.[9], which were calculated with the MTP machine learning potential, showing a considerable difference with the thermal conductivity of the present study and also with Ref. [12]. The reason for this discrepancy could be that the potential used in Ref. [9] was not properly trained or the DFT input dataset was not complete. Also, Young's modulus was determined as 115.7 N/m through present non-equilibrium simulations, corresponding to 368 GPa for the thickness analyzed in this study, which amounts to 3.18 Å.

Table 1: Thermal conductivity and elastic properties of Penta-NiN$_2$ at T = 300 K.

| Method | $k$ (W/mK) | Elastic properties (N/m) | Ref. |
| --- | --- | --- | --- |
| DLP | 133 | E=115.7 | Present study |
| DFT [a] | 109.2 | $C_{11}= C_{22}$=172.2  $C_{12}$=21.3, $C_{66}$=48 | [12] |
| DFT | - | E=168.8 | [11] |
| MTP [b] | 610 | E=194.3 | [9] |
| DFT | - | $E_{max}$= 231.2  $E_{min}$=14.9 | [10] |

[a] The considered thickness is 6.80 Å
[b] The considered thickness is 2.85 Å

**Conclusion**

In this study, we employed a deep learning-based interatomic potential to investigate the thermal and mechanical properties of single-layer penta-NiN$_2$, a promising 2D material with diverse applications. Key highlights of the research include:

- *Size-Dependent Thermal Conductivity:* The study revealed that penta-NiN$_2$'s thermal conductivity increased from low values at shorter lengths to a converged value around 133 W/mK at infinite length, pointing to a diffusive-like regime in this limit.

- *Temperature-Dependent Behavior:* Through a broad temperature range, we demonstrated that penta-NiN$_2$'s thermal conductivity exhibited an inverse proportionality to temperature (k∼T$^{-1}$).
- *Strain Effects:* The impact of tensile strain on penta-NiN$_2$'s thermal conductivity was explored, assessed that increasing strain levels led to a reduction in thermal conductivity due to decreased phonon group velocity.
- *Mechanical Response:* Penta-NiN$_2$'s mechanical properties were evaluated under various loading conditions, yielding Young's modulus of approximately 368 GPa, consistent with existing literature. The material exhibited robust structural stability. Also, the potential that was trained without considering strain proved incapable of preserving structural stability under high strain levels.
- *Defect Sensitivity:* The study highlighted the significant influence of vacancy defects on the material's mechanical strength, with a 2% defect concentration resulting in notable reductions of 38% in Young's modulus, 29% in fractural strain, and 45% in ultimate strength compared to defect-free systems.
- *Deep Learning Interatomic Potentials:* This research showcased the effectiveness of deep learning-based interatomic potentials in predicting both thermal and mechanical properties, offering a powerful and reliable tool for materials research.

In conclusion, this study advances our understanding of penta-NiN$_2$'s thermal and mechanical behavior, providing quantitative insights into its potential applications in nanoelectronics and beyond. The research also underscored the promise of deep learning interatomic potentials in the systematic investigation of 2D materials, offering a data-driven approach to material characterization and design.

**Data availability**

The data that support the findings of this study are available upon reasonable request. Please contact the corresponding author for data inquiries.